# Nonthermal effects in solids after swift heavy ion impact

N. Medvedev[1,2,*], R. Voronkov[3], A.E. Volkov[3]

*1) Institute of Physics, Czech Academy of Sciences, Na Slovance 2, 182 21 Prague 8, Czech Republic*
*2) Institute of Plasma Physics, Czech Academy of Sciences, Za Slovankou 3, 182 00 Prague 8, Czech Republic*
*3) P.N. Lebedev Physical Institute of the Russian Academy of Sciences, Leninskij pr., 53,119991 Moscow, Russia*

## Abstract

This contribution is a brief introduction to nonthermal effects related to modifications of the interatomic potential upon ultrafast excitation of the electronic system of solids, primarily focusing on the swift heavy ion track problem. We clarify the difference between the exchange of the *kinetic* energy of electrons (and holes) scattering on the lattice (electron-phonon coupling, "thermal effects") and the relaxation of the nonequilibrium *potential* energy of a solid ("nonthermal effects"). We discuss that at different degrees of electronic excitation, the modification of the interatomic potential may result in various phase transitions without an increase of the atomic temperature, i.e., at room temperature (nonthermal melting, formation of the superionic state), or in atomic acceleration causing "nonthermal heating" of the target atoms. Examples of theoretically predicted various effects are given, supported by known experimental observations.

## I.    Introduction

When a swift heavy ion (SHI) impinges on a solid in the electronic stopping regime ($E > 1$ MeV/nucl., $S_e >> S_n$, where $S_e$ and $S_n$ are the electronic and nuclear energy losses of the ion, respectively), it deposits a large amount of energy into the electronic system of the target [1,2]. Spreading outwards from the ion trajectory, excited so-called delta-electrons then redistribute this energy *via* electronic cascades exciting secondary electrons, and creating deep-shell and valence holes [3]. The electronic kinetics eventually leads to energy transfer to the lattice. Atomic response to this energy transfer results in nanometric material modifications along the ion trajectory known as the SHI track [1,2].

High gradients of the initially deposited energy result in fast spatial spreading of the created electronic excitation. This causes a quick decrease of the electronic energy density and restricts the time of energy transfer to the atomic system in the proximity of the ion trajectory by ~100 fs after the ion impact. During this time, the state of the electronic system is evolving fast, governed by the transport of electrons and holes and by energy exchange with the target atoms.

Since the 1960-s, it has been assumed that the electron energy transfer to the atoms may be described in terms of the so-called two temperature model [4,5] (a later modification of which became known as the inelastic thermal spike model, i-TS [6]). This model describes the energy exchange between the electronic and the atomic system as the scattering process, often presumed to be electron-phonon scattering. Since the 1990-s, the i-TS model dominated the field of SHI track formation, although the key parameter of the model, the electron-phonon coupling factor, is a fitting parameter in the model. It is fitted so that the i-TS calculated SHI track radius, defined as the distance at which the lattice temperature reaches the macroscopic melting temperature of the material, coincides with the radius of a damaged region measured in experiments [6,7]. When fitted in this way, it results in extremely high values of the coupling parameter (~$10^{18}$-$10^{19}$ W/(m$^3$K)) [7,8], originated from the fact that atoms must heat up to the melting temperature before electrons cool down (~100 fs).

---

*Corresponding author: nikita.medvedev@fzu.cz

At the same time, the electron-phonon coupling parameter is accessible in laser pump-probe experiments, e.g., *via* time-resolved diffraction measurements that give access to the atomic temperature evolution [9,10]. This kind of experiment, supported by various theoretical calculations [9,11–14], demonstrated that the electron-phonon coupling is a relatively slow process, requiring a picosecond timescale for equilibration between the electronic and the atomic temperatures. The measured and calculated electron-phonon coupling parameters are much smaller (~$10^{16}$-$10^{17}$ W/(m$^3$K)) than that extracted from SHI tracks data with the fitting of the i-TS (see Figure 1 in Ref. [8]).

Recently, this conundrum was resolved by noticing that the electron-phonon coupling is not the only mechanism of the energy transfer in matter from a highly excited electronic system to the atomic one [8]. As electrons form the attractive part of the interatomic potential, which keeps atoms of a solid together under normal conditions, the (extreme) excitation of the electronic system directly affects the interatomic potential. It is known from laser experiments that in some materials, electronic excitation may lead to ultrafast atomic lattice destabilization even at room atomic temperature – hence, it was named "nonthermal melting" [15–18]. It may be regarded as the breaking of interatomic bonds induced by the electronic excitation instead of the increase of the atomic temperature. Changing the atomic potential can also result in the acceleration of atoms causing an increase in their kinetic energy in the proximity of the ion trajectory. In contrast to *kinetic* energy (momentum) exchange between the electronic and atomic system during scatterings acts, this process manifests a conversion of the excited *potential* energy of material into the kinetic energy (heating) of the atomic system.

A possibility of heating of atoms by mechanisms other than electron-phonon coupling was previously discussed in the literature using the Coulomb explosion model [19]. This model assumes transient charge separation in the proximity of the SHI trajectory: delta-electrons transport leaves positively-charged track core. Positive ions of the target start to repeal each other via the Coulomb, thus accelerating. The major shortcoming of that idea was that the Coulomb explosion is only known to take place in finite-size systems: small molecules and nanoclusters, or possibly thin layers or at the surface of materials, but not in SHI tracks in solids [20]. Unlike in gases/plasma, the positive charge in solids is not bound to parent ions – valence holes may quickly travel out of the track similarly to conduction-band electrons. At the same time, a large number of fast electrons generated by the SHI neutralizes the uncompensated charge within a few femtoseconds, as was measured experimentally [21,22]. This time is too short for atoms to gain significant energy. In contrast, nonthermal effects discussed here do not require an unbalanced charge. From the theoretical viewpoint, Coulomb explosion is a particular case of the nonthermal effects, in which the electrons are ionized and fly far enough from the parental ion, instead of an excitation preserving charge neutrality – meaning, the same *ab-initio* methods could be used to describe both effects.

The recent works revealed that there are three major channels of energy exchange between the electronic and the atomic systems of a solid in the proximity of the SHI trajectory: (a) quasi-elastic scattering of excited electrons (scattering on the lattice delivering kinetic energy to atoms, without secondary electron excitation; studied in detail, e.g., in Refs. [3,23]), (b) quasi-elastic scattering of valence holes (see Ref. [24]), and (c) the nonthermal modifications of the interatomic potential, resulting in atomic acceleration (described in Ref. [8]). Combining these three channels of the atomic energy increase results in atomic heating sufficient to create observable SHI tracks [25–27].

In this brief review, we discuss the nonthermal effects in various materials. We clarify the mechanisms of modification of the interatomic potential and discuss the levels of electronic excitation that are necessary for the ultrafast atomic disorder. We also discuss the fundamental difference between thermal effects and nonthermal ones. We provide some examples of various manifestations of the nonthermal effects, under some conditions resulting in the production of exotic material states. Finally, we discuss the application of a model, accounting for the three channels of energy exchange, to an SHI track formation problem, which enables the calculation of the track formation thresholds in agreement with experimental data without the need for fitting parameters [26,27].

## II.  Thermal and non-thermal effects

Let us start by describing in more detail the terms used throughout the paper.

*Thermal effects.* Generally, in the course of the exchange of the kinetic energy between excited electrons and atoms, no deformation of the interatomic potential by electronic excitations is assumed. Furthermore, it is usually supposed that the atomic system may be described as an ensemble of phonons – quasiparticles of atomic vibrations around their unchanged equilibrium positions (see the illustration in Figure 1a). Due to a large difference between the masses of atoms and electrons, after a number of scattering acts atoms (a) remain near their equilibrium positions and (b) gradually increase their kinetic energy. Additionally, assuming fast thermalization in the atomic ensemble, this process of the electrons-atoms energy exchange may be treated as a heating of the atoms. Phase transformations occur when the atomic temperature reaches the necessary threshold values.

The key point here is that the thermal effects are the result of the exchange of the kinetic energy between electrons and atoms of a solid governed by electron-atoms scattering. Because the scattering involves a transfer of momentum, or the atomic translation operator, one may say that thermal heating is the energy transfer between electrons and the atomic movements (e.g., vibrations – phonons).

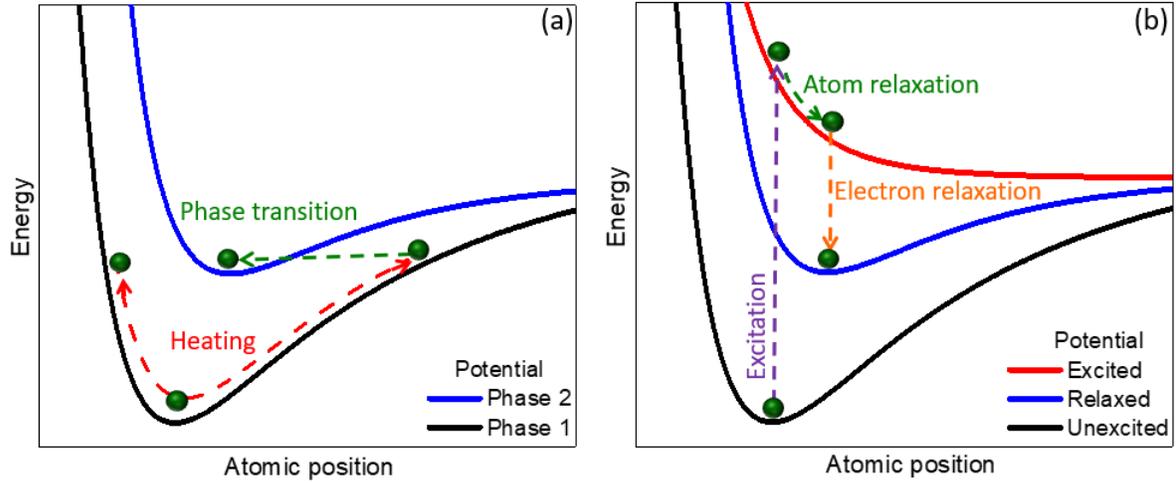

*Figure 1. Illustration of thermal (a) and nonthermal (b) processes.*

Hence, in the field of ultrafast energy deposition in solids (such as an SHI or ultrashort laser pulses irradiation), the electron-phonon coupling is the parameter defining the thermal effects. Strictly speaking, the definition of a phonon assumes a crystalline lattice, harmonic atomic vibrations, and timescales longer than the atomic oscillation period (inverse phonon frequency) – the conditions that are not fulfilled in the ion irradiation scenario – but this point is beyond the scope of this paper and we refer an interested reader to the earlier discussion of this point in Refs. [14,28]. In the field of ultrafast material science, the term "coupling" refers to the electron-ion (electron-phonon) scattering.

The scattering is a non-adiabatic (non-Born-Oppenheimer) process [29,30]. The joint wave function of the system cannot be represented as a simple product of the electronic and atomic wave functions during a scattering act.

*Nonthermal effects*. These effects originate from the atomic response to the changes in the interatomic potential due to an excitation of the electronic ensemble, see Figure 1b.

The overlap of the electronic wave functions (orbitals) defines the atomic potential energy surface of a solid [31,32]. When electrons are excited, they change their states, e.g., they are promoted from the valence band into the conduction band. The electronic system in this new state forms new spatial distribution of the electron density changing interaction of neighboring atoms compared to their initial distribution. This means that atomic ensemble instantly starts to experience different forces than in the equilibrium state, which may initiate atomic rearrangement.

The key point is that the nonthermal effects are the response of atoms to the changes in the nonequilibrium potential. These changes may cause phase transformations or a transfer of the excited potential energy of a solid to the kinetic energy (heating) of the atomic ensemble.

A schematic illustration of modifications of the interatomic potential is shown in Figure 2 for different levels of the electronic excitation. At low levels of electronic excitation, a small perturbation of the interatomic potential leads to an atomic reaction, known as displacive excitation of coherent phonons [33]. It means, atoms receive a coherent kick inducing some new vibrations. Such a change in the potential may result in a slight increase or decrease in the atomic kinetic energy [34].

At a higher electronic excitation level, reaching a threshold for a phase transition, the material ends up in a different phase (e.g., another solid phase, or a liquid). Figure 2 illustrates that no thermal gradual acceleration of atoms occurs to levels triggering a structure or phase transformation.

At extremely high excitation levels, the interatomic potential may locally turn from an attractive to a repulsive one, accelerating atoms vastly.

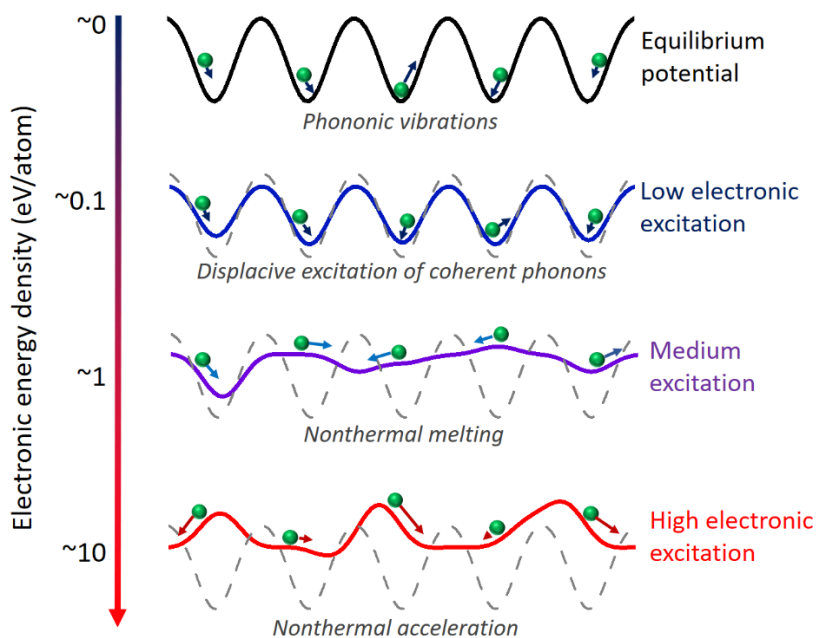

*Figure 2. Schematic illustration of the interatomic potential changes with an increase of the energy deposited into the electronic system of a solid (increasing electronic temperature). The upper panel shows atoms (green balls) oscillating in an unexcited equilibrium potential (black line). The middle panel shows flattened excited potential (purple line), resulting in nonthermal melting (at the room atomic temperature); the former equilibrium potential is depicted with the dashed grey line for comparison. The lower panel shows a highly-excited potential turning repulsive (red line), thus accelerating atoms resulting in "nonthermal heating" of atoms.*

Light electrons are typically so much faster than ions that one may assume that they instantly adjust to any atomic displacement. This is known as the Born-Oppenheimer approximation (BO), describing adiabatic changes of the electron-ion Hamiltonian during atomic motion. In contrast to the scattering, no coupling between the electronic and atomic wave functions appears in this approximation [34,35]. That means, within BO there is no coupling to the atomic motion, including atomic vibrations (electron-phonon coupling), thus this approximation cannot capture thermal effects. As a result, the electronic and atomic temperatures cannot equilibrate within the BO approximation [36].

In the framework of the BO approximation, electrons follow the guiding dynamics of ions and instantaneously adjust to the changes in the band energy levels induced by atomic motion [29]. The adiabatic approximation means no changes in the electronic populations (the distribution function) on the evolving band energy levels [34,35]. The unchanging electronic distribution function results in no electron transitions between these levels and, again, no exchanges of the kinetic energy of electrons with the atomic system [29,30].

Standard *ab-initio* simulation technics such as the density functional theory (DFT), or its simplified version, the tight-binding (TB) method, combined with the molecular dynamics tracing atomic trajectories, are well equipped to trace nonthermal effects. In this work, we use the TB code XTANT-3 [8,37], and the DFT package Quantum Espresso [38] to illustrate the nonthermal effects in solids.

To model thermal effects, one needs to employ so-called *non-adiabatic* methods such as, e.g., time-dependent DFT or other technics [39–41]. Non-adiabatic methods account for the fact that electrons react to an atomic motion not instantaneously when they change their state and their density around ions. It results in kinetic energy exchange between electrons and ions/atoms. These methods are beyond the scope of the current paper but may be found, e.g., in Refs. [11,41].

### II.1. Nonthermal melting: disorder without atomic heating

Material damage after excitation of its electronic system is typically a threshold process, it takes place only above a certain energy density deposited, a particular value of which depends on the properties of the material and the radiation. Near the threshold deposited dose (or threshold electronic temperature) for nonthermal effects, the interatomic potential turns rather flat (see a schematic illustration in Figure 2). Under such conditions, even the inertial movement of atoms at room temperature is sufficient for the material disorder [42]. Thus, the material melts at room temperature in the atomic system [17,43]. For that reason, early works referred to this effect as "cold melting" [43,44]. However, it was quickly realized that this term is misleading, since the electronic temperature in this process must be very high, and the material as a whole is not cold. Nowadays, the commonly accepted term is "nonthermal melting" (or, more generally, nonthermal phase transition), which signifies that the melting of the material is not induced by thermal means (an increase of the kinetic energy of atoms), but by a different mechanism: flattening of the atomic potential energy surface due to a significant electronic excitation.

Under normal conditions, without electronic excitation, the material's melting temperature is defined as the average energy required by atoms to overcome their potential barriers and disorder. In an electronically-excited state, the potential barriers are lowered or may completely disappear (see the illustration in Figure 2). In the process of nonthermal melting, instead of rising the atomic temperature to the melting point, the reverse happens: the melting point of the material lowers down to a below-room-temperature value.

Initially, nonthermal melting was theoretically predicted [45–47] and experimentally demonstrated for ultrafast-laser-irradiated covalently bonded semiconductors [15,17,18,48]. It is now a well-established effect in the ultrafast science community. Later, it was predicted to take place in insulators, including oxides [49,50], ionic solids [51,52], and polymers [53,54]. It appears to be a universal effect in crystalline non-metallic materials. In metals, nonthermal effects manifest themselves in a more complex manner, as will be discussed below.

### II.2. "Nonthermal heating": acceleration of atoms due to changes in interatomic potential

At deposited energy densities significantly above the damage threshold, the modifications of the interatomic potential may be so violent that the atoms may significantly accelerate in response (see schematics in Figure 2) [8]. The acceleration of atoms is due to modification of the interatomic potential, and not as a result of the equilibration of the electronic and atomic temperatures. In this sense, this acceleration may be regarded as "nonthermal heating" of atoms: an increase of the kinetic energy of atoms *via* non-thermal means. This effect is fundamentally different from electron scattering exchanging the *kinetic* energy between the electrons and atoms (electron-phonon coupling). It is a conversion of the modified *potential* energy into the kinetic one, which is not limited by the electron-phonon coupling rate.

Atomic acceleration as a result of the modification of the interatomic potential was experimentally observed with the help of ultrafast diffraction in bismuth [44] and silicon [42].

To induce such a nonthermal acceleration of atoms, it requires a significant deposited energy density, on the order of a few eV/atom or higher. Such doses are easily reached in an SHI irradiation scenario but occur in laser spots only at high fluences, e.g. in laser ablation experiments. Since most of the experiments measuring electron-phonon coupling are performed at much lower doses, they have access to the slow electron-phonon coupling process without the interference of the fast nonthermal effects.

The nonthermal atomic acceleration and the electron-phonon coupling are independent processes, but the two may influence each other. The electron-ion (electron-phonon) coupling parameter is a multivariable function of the electronic and the atomic temperatures, atomic structure, density, etc. [11]. Its dependence on the atomic temperature is approximately linear [11]. Ultrafast nonthermal heating, increasing the kinetic energy of atoms, also increases the electron-ion coupling. Therefore, this enhanced coupling raises the atomic temperature even further. This synergy results in a very fast increase in the atomic temperature [8].

Atomic movement during nonthermal heating can cause drastic changes in the electronic structure (band structure) of the material. In the case of covalent insulators and semiconductors, typically the bandgap of the material collapses, producing metallic-like properties, see an example of alumina in Figure 3 [36,37,55]. The energy, associated with the band gap of the material, is released into the kinetic energy of atoms. Note that the initial bandgap energy in $Al_2O_3$ in Figure 3 is $E_{gap}(t<0$ fs$)\sim$9 eV, whereas shortly after the electronic excitation with the dose of 5 eV/atom, the bandgap shrinks to $E_{gap}(t>100$ fs$)\sim$0 eV – this energy difference is converted into the kinetic energy of atoms [8]. Figure 3 demonstrates that nonthermal heating of atoms due to changes of the interatomic potential may take place within $\sim$100 fs, the time comparable to the life-time of the short-lived electronic excitation in SHI tracks [8].

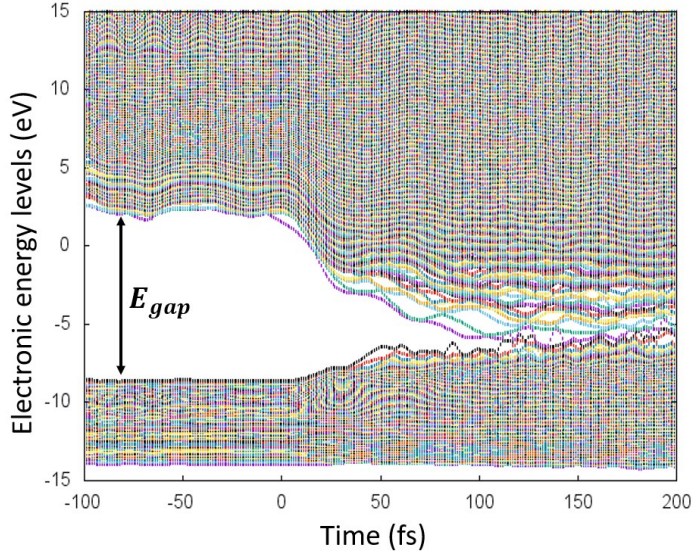

*Figure 3. An example of the evolution of the electronic energy levels in Al₂O₃ after the deposition of the dose of 5 eV/atom into its electronic system at the time t=0 fs. Calculations performed with the tight-binding molecular dynamics code XTANT-3 [37]. The initial (unexcited) material bandgap is illustrated with the arrow.*

Estimates of the "electron-phonon coupling" extracted from the swift heavy ion tracks parameters based on the i-TS model must be interpreted as reflecting mostly the rate of nonthermal increase of the atomic kinetic energy (with a contribution of the electron- and valence hole-atom scattering) – not as a real electron-phonon coupling parameter [8]. The connection between the bandgap collapse and the atomic heating helps to explain the empirical dependence observed in the i-TS model: the "coupling parameter" seems to be larger in materials with larger bandgaps [7]. This dependence must be interpreted as the observation of faster atomic heating in large bandgap materials, which is a natural outcome of nonthermal heating [8]. Indeed, since atoms accelerate as a result of the bandgap collapse, they receive more energy in wide-bandgap materials. This consideration has one caveat: not in all materials the bandgap collapses completely. That means the dependence between the speed of atomic heating and the bandgap value will have exceptions, which was also noticed empirically [7]. This point will be discussed in more detail in the next section.

Note that in all considerations above, we did not assume the presence of an uncompensated charge. All the processes discussed take place even within a charge-neutral system. In an SHI track, there may also be a transient charge non-neutrality: e.g. during the stage of ballistic transport of electrons, or long-lasting charge non-neutrality in finite-size systems (such as molecules and nano-clusters), or in near-surface regions of solids, where emitted electrons may not return. In this case, all the above-mentioned phenomena would still take place, but there will be additional processes occurring associated with the Coulomb explosion due to the uncompensated charge [56,57]. That would lead to an even stronger and faster acceleration of atoms, thereby reinforcing our conclusions.

### III. Examples of nonthermal effects

#### *III.1.     Nonthermal melting in covalent materials*

Nonthermal melting in semiconductors was experimentally studied for a long time, including such materials as silicon [58], germanium [17], GaAs [59], and other group III-V compounds [60]. Those materials were also investigated theoretically [45,46,61–63]. They all exhibit a very similar response to high levels of electronic excitations: the bandgap collapses, and atoms disorder within ultrashort timescales, on the order of a few hundred femtoseconds at near-threshold deposited doses. The final material phase after ultrafast nonthermal phase transition may be different, depending on the particular material: while most of the materials disorder into a regular liquid phase, silicon, and AlAs were predicted to have two phases – (transient) low-density and high-density liquids [36,63].

The response of wide-bandgap covalent materials to high electronic excitation so far was only studied theoretically in such materials as diamond [37,64], various oxides [49], and polymers [53,54]. Those materials are predicted to also exhibit very similar behavior initiated by a bandgap collapse. An example of the bandgap collapse and respective rise in the atomic temperature in SiO$_2$ are shown in Figure 4. The figure also illustrates atomic snapshots showing progressing disorder induced at ultrashort timescales. It is clear that with the collapse of the bandgap, atomic temperature increases, reaching its maximum when interatomic bonds start breaking and a phase transition ensues. The simulation was performed within BO approximation, which excludes electron-phonon coupling, thus clearly demonstrating the nonthermal heating effect [8].

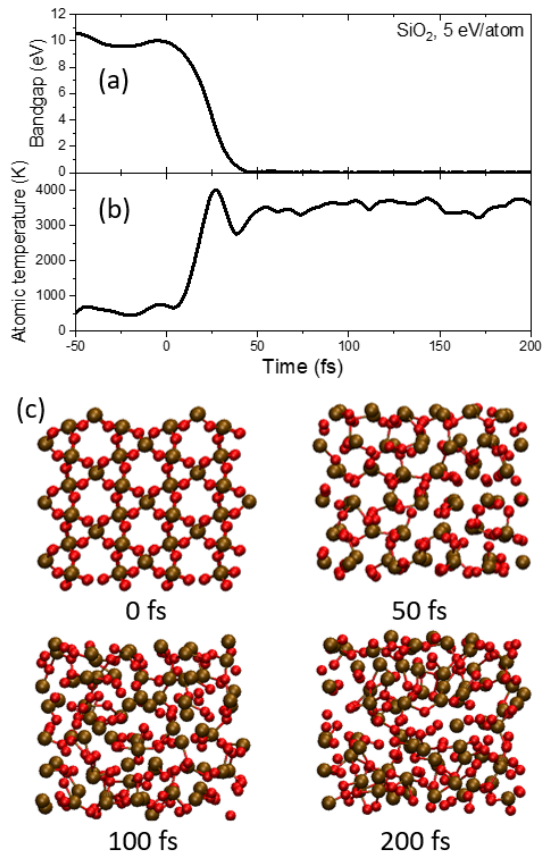

*Figure 4. Simulated quartz irradiated with the deposited dose of 5 eV/atom into its electronic system at t=0 fs. (a) Bandgap as a function of time. (b) Evolution of the atomic temperature. (c) Atomic snapshots at different times after irradiation. Simulations were performed with help of XTANT-3 code* [49]. *Red balls are oxygen atoms, large orange balls are silicon.*

### III.2. Nonthermal melting in ionic crystals

Similar to covalent compounds, ionic crystals may exhibit different types of nonthermal transformations such as amorphization or melting depending on the material [51]. For example, our simulations performed with the *ab-initio* package Quantum Espresso demonstrate that NaCl crystal at electronic excitation doses exceeding ~2.5 eV/atom turns into an amorphous solid – a phase that is hardly achievable with usual methods [65].

However, in contrast to covalent materials, band gaps of ionic compounds with high bond ionicity (such as alkali-halides, Pauling bond ionicity $f \geq 0.9$ [66]) are very stable. They may shrink by a few eV but do not collapse even at very high deposited doses. Typical band gap behavior of ionic crystals during nonthermal transitions at the electronic temperature $T_e = 6$ eV is shown in Figure 5 on the example of three compounds: LiF, KBr and NaCl [51].

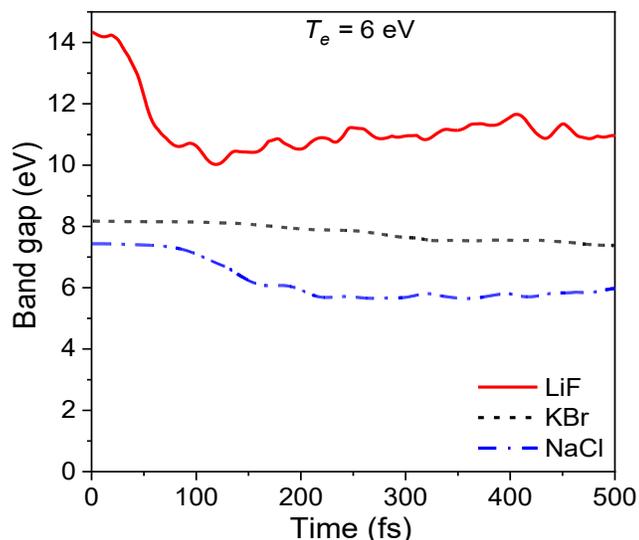

*Figure 5. Evolution of band gaps of LiF, KBr, and NaCl during nonthermal transitions at electronic temperature $T_e = 6$ eV corresponding to deposited doses: 10.5 eV/atom for LiF, 19.8 eV/atom for NaCl, 21.9 eV/atom for KBr.*

Materials with an intermediate bond ionicity such as $Al_2O_3$ (Zhuravlev bond ionicity $f = 0.69$ [67]) or $Y_2O_3$ (Zhuravlev bond ionicity $f = 0.71$ [67]) also demonstrate intermediate band gap stability: it may collapse but not as easily as in covalent materials. In such compounds, band gap collapse requires higher deposited doses than those necessary to trigger nonthermal transitions at subpicosecond timescales – in contrast to covalently bonded materials (Pauling bond ionicity $f \leq 0.4$ [66]) where thresholds triggering nonthermal transformations and band gap collapse coincide [51,68].

### III.3.     Nonthermal effects in metals

Bulk metals after high electronic excitation typically exhibit the opposite effect: phonon hardening [69,70]. That means the interatomic potential becomes stiffer instead of softer or flatter. However, some bcc metals may also have softening of the potential, depending on the level of electronic excitation [71,72]. On top of this, another complication with metals is that in finite-size samples, an increase in the electronic temperature leads to an increase in the electronic pressure. That, in turn, leads to material expansion. The expanded lattice may then destabilize and disorder, without significant heating *via* electron-phonon coupling [73]. So, the finite-size metals have a different kinetic pathway to nonthermal melting – *via* intermediate expansion due to the electronic pressure. Since metals typically do not form SHI tracks, we will not go into details of nonthermal effects in metals any deeper here.

### III.4.     Formation of unusual material phases

Electronic excitations in condensed matter form unique conditions that bring material out of the conventional pressure-temperature phase diagram into a 3-dimensional one: pressure, ionic temperature, electronic temperature. Points on this diagram may correspond to phases that are unachievable for material by conventional methods. Such out-of-equilibrium state of matter is known as the two-temperature state [74–76] – not to be confused with the two-temperature model as one of many possible methods describing this state.

One of the most exotic examples of such phases is the superionic state, i.e., a solid with very high ionic conductivity. Originally, this name referred to defective crystals in which ions may demonstrate liquid-like diffusion via vacancies and interstitials [77]. However, it was discovered that similar values of ionic conductivity may be obtained in pure crystals by applying high pressure and temperature to water and water-based mixtures [78–80]. Under such conditions, superionic water (also known as ice-XVIII) is formed from the solid oxygen-ion lattice with hydrogen ions (protons) moving through this lattice as a liquid.

Recent DFT simulations revealed that a similar state may be obtained upon electronic excitation in group III oxides: with liquid oxygen sublattice and solid metallic one [50]. Figure 6 demonstrates typical mean atomic displacements for a nonthermal phase transition to the superionic state on the example of $Al_2O_3$. One can see that oxygen atoms' displacements are proportional to the square root of time which corresponds to diffusive (liquid) behavior, whereas aluminum atoms oscillate around equilibrium positions indicating a stable crystalline lattice [68].

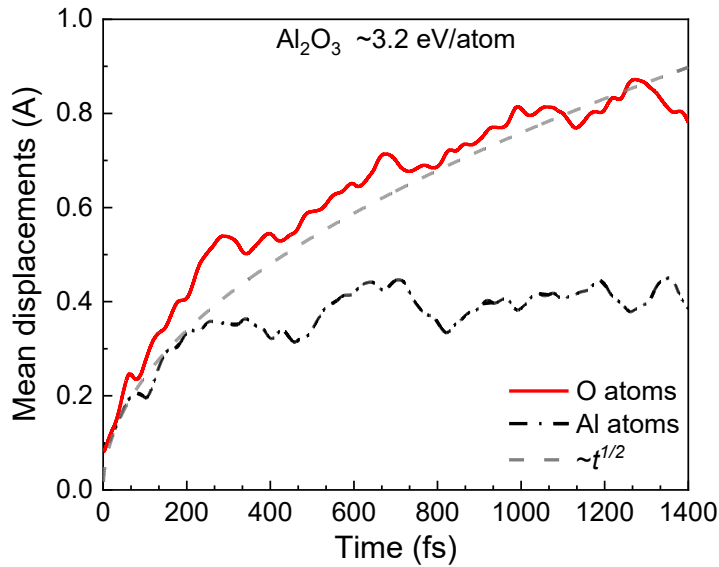

*Figure 6. Mean atomic displacements of $Al_2O_3$ after the deposited dose of ~3.2 eV/atom.*

Superionic states are produced in a range of deposited doses up to a certain threshold at which metallic sublattice also melts. At doses between the superionic and melting thresholds the band gap collapse may occur within the superionic phase. That opens a prospect for the fabrication of superionic states with different levels of electronic conductivity [68].

Similar superionic-like states were also predicted to form in irradiated polyethylene [53]. In this polymer, after excitation of the electronic system to the doses of 0.05 eV/atom, hydrogen atoms start to detach from the carbon backbone. At doses of ~0.3-0.5 eV/atom, the entire hydrogen subsystem behaves as a liquid, whereas the carbon subsystem is still intact.

### III.5.     A simple model to account for nonthermal heating in SHI tracks

Knowing the importance of nonthermal heating, it is clear that this channel of energy transfer from electrons to atoms in SHI tracks must be accounted for in reliable simulations. All the three most important channels of the energy exchange may be traced with kinetics methods, such as Monte Carlo simulation: quasi-elastic scattering of electrons, valence holes, and nonthermal effects.

As was discussed above, in covalent materials, the nonthermal effect manifests itself as the bandgap collapse. Thus, one can build a simple model of nonthermal atomic heating, simply by tracing the distribution in space of the electron-hole pairs formed in cascades induced by an SHI [25]. The value of the energy transferred nonthermally (*via* band gap collapse) may be estimated simply as the bandgap of the material times the number of the electron-hole pairs [25–27]. In other than covalent materials, with incomplete bandgap collapse, further dedicated studies are still required.

The energy transferred *via* the three channels may be fed to atoms modeled with the classical molecular dynamics (MD) method. Since by the time of ~100 fs electrons in an SHI track already dissipate outwards from the track core, the interatomic potential returns to its unexcited form [25]. After that, standard MD simulation becomes applicable.

As was demonstrated in Ref. [25], to reproduce the *final* material modification in a track, this initial stage of ~100 fs when electronic kinetics develops in parallel with the atomic one, does not require a simultaneous tracing of both systems: the atomic dynamics may be started to be simulated with MD after the 100 fs, using the energy transferred from electrons as initial conditions. This significantly simplifies the modeling, since Monte Carlo simulations of the electronic processes may then be decoupled from the MD simulation of the atomic response.

When profiles of the energy transfer from the quasi-elastic scattering of electrons, holes, and nonthermal bandgap collapse, are fed to atoms as initial conditions, the atoms then react and may disorder (at sufficiently high stopping powers of an SHI) [27].

During cooling of the formed track, typically requiring a few hundred picoseconds, the final modifications may be formed[81]. A comparison of thusly calculated track thresholds with experimentally measured ones in $Al_2O_3$ demonstrated a reasonable agreement [26]. It validated the understanding of the mechanisms of atomic heating in SHI tracks. Such simulations require no *a posteriori* fitting parameter. That makes such models fully predictive.

For those simulations, the Monte Carlo code TREKIS-3 was used to prepare the initial conditions for further LAMMPS molecular dynamics simulations [82]. TREKIS-3 is publicly available at https://github.com/N-Medvedev/TREKIS-3.

## IV. Conclusions

Nonthermal atomic heating and structure transformations occur due to a modification of the interatomic potential induced by extreme electronic excitation. Conversion of the nonequilibrium potential energy of the excited solid into the kinetic energy of the atomic ensemble, supplemented by electron and hole scattering, drives the atomic kinetics in SHI tracks.

The nonthermal effects are present in all materials but may manifest differently depending on the class of solid and the level of excitation. In wide-bandgap covalent materials, the effect is most pronounced at doses above the threshold for nonthermal phase transition: the wide bandgap collapsing within an ultrashort timescale (~100 fs) provides the highest rate of the energy exchange from electrons to atoms resulting in fast acceleration of atoms ("nonthermal heating"). The interatomic potential at the same time is transiently drastically changed, reducing the barrier for the disorder, thereby inducing ultrafast nonthermal phase transitions.

Identifying the most important mechanisms of energy transfer from the SHI-excited electrons to atoms of the target enables one to build reliable models. Combining the three channels: quasi-elastic scattering of excited electrons, valence holes, and nonthermal heating, in a simple model for covalent materials, enabled us to describe SHI track formation thresholds in good agreement with experimental data avoiding the usage of fitting parameters. It validated the model, confirming the importance of the nonthermal channel of energy exchange.

## Acknowledgments

Computational resources were supplied by the project "e-Infrastruktura CZ" (e-INFRA LM2018140) provided within the program Projects of Large Research, Development and Innovations Infrastructures. This work has been carried out in part using computing resources of the Federal collective usage center Complex for Simulation and Data Processing for Mega-science Facilities at NRC "Kurchatov Institute", http://ckp.nrcki.ru/. NM gratefully acknowledges financial support from the Czech Ministry of Education, Youth and Sports (grants No. LTT17015, LM2018114, and No. EF16_013/0001552). This work benefited from networking activities carried out within the EU funded COST Action CA17126 (TUMIEE) and represents a contribution to it. AEV and RV acknowledge support from the Russian Science Foundation (grant number №22-22-00676, https://rscf.ru/en/project/22-22-00676/).